\documentclass[conference]{IEEEtran}
\IEEEoverridecommandlockouts
\usepackage{makecell}
\usepackage{cite}
\usepackage{amsmath,amssymb,amsfonts}
\usepackage{algorithmic}
\usepackage{graphicx}
\usepackage{textcomp}
\usepackage{xcolor}
\usepackage{multirow}
\usepackage{pifont}
\usepackage{marvosym}
\usepackage{balance}
\usepackage{amsmath}
\usepackage{hyperref}

\def\BibTeX{{\rm B\kern-.05em{\sc i\kern-.025em b}\kern-.08em
    T\kern-.1667em\lower.7ex\hbox{E}\kern-.125emX}}
\begin{document}

\title{DCIM-AVSR:Efficient Audio-Visual Speech Recognition via Dual Conformer Interaction Module}
\author{
    \IEEEauthorblockN{Xinyu Wang$^{1}$, Haotian Jiang$^{1}$, Haolin Huang$^{1}$, Yu Fang$^{1}$, Mengjie Xu$^{1}$, Qian Wang$^{1}$$^{\ast}$}
    \IEEEauthorblockA{\textit{$^{1}$School of Biomedical Engineering \& State Key Laboratory of Advanced Medical Materials and Devices} \\
    \textit{ShanghaiTech University},
    Shanghai, China
    }
}

\maketitle

\renewcommand{\thefootnote}{}
\renewcommand{\footnoterule}{%
    \kern -3pt
    \hrule width 0.3\textwidth height 0.6pt 
    \kern 2pt
}
\footnotetext{$\ast$ wangqian2@shanghaitech.edu.cn} 


\begin{abstract}
Speech recognition is the technology that enables machines to interpret and process human speech, converting spoken language into text or commands. This technology is essential for applications such as virtual assistants, transcription services, and communication tools. The Audio-Visual Speech Recognition (AVSR) model enhances traditional speech recognition, particularly in noisy environments, by incorporating visual modalities like lip movements and facial expressions. While traditional AVSR models trained on large-scale datasets with numerous parameters can achieve remarkable accuracy, often surpassing human performance, they also come with high training costs and deployment challenges. To address these issues, we introduce an efficient AVSR model that reduces the number of parameters through the integration of a Dual Conformer Interaction Module (DCIM). 
In addition, we propose a pre-training method that optimizes model performance by fine-tuning.
Unlike conventional models that require the system to independently learn the hierarchical relationship between audio and visual modalities, our approach incorporates this distinction directly into the model architecture. This design enhances both efficiency and performance, resulting in a more practical and effective solution for AVSR tasks.

\end{abstract}


\section{Introduction}

In recent years, automatic speech recognition (ASR)~\cite{yao21_interspeech, baevski2020wav2vec, radford2023robust,ng23binterspeech,xiao22interspeech,wild,xiaoicassp} has rapidly advanced, driven by deep learning and end-to-end neural approaches. However, ASR remains challenging in complex acoustic environments, such as overlapping speech, noise, and reverberation. To address this, researchers are increasingly incorporating visual features, like lip movements and facial expressions~\cite{yeo2024visual, kim2021cromm}, into ASR models. This integration, known as audio-visual speech recognition (AVSR)~\cite{xia2020audiovisual}, helps reduce the impact of distorted speech signals.

Recent AVSR works have introduced various methods to enhance recognition ability~\cite{burchi2024multilingual, ma2023auto, chang2024Conformer, DBLP:conf/iclr/ShiHLM22, shi22_interspeech}. For example, Auto-AVSR~\cite{ma2023auto} used 12 layers of Conformer~\cite{gulati20_interspeech} for processing both visual and audio data. In contrast, Fast Conformer~\cite{burchi2024multilingual} used 18 layers, with the first 10 focused separately on visual and audio processing and the final 8 layers acting as a combined decoder. LP Conformer~\cite{chang2024Conformer} emphasized the visual front-end, exploring different visual architectures. While these approaches achieve state-of-the-art results, they require significant training resources.

Computational efficiency is crucial in AVSR research but is often overshadowed by the focus on performance improvements. Many end-to-end AVSR models use an audio-visual bi-encoder framework, which demands large datasets and complex models. To solve this problem, some researchers are beginning to explore more efficient AVSR approaches. HOURGLASS-AVSR~\cite{yu2024hourglass} proposed an hourglass AVSR model that reduces computational complexity by shortening the time dimension of intermediate features and performing multi-modal alignment, thereby achieving both high efficiency and performance. MLCA-AVSR~\cite{wang2024mlca} proposed a multi-layer cross-attention fusion module to achieve more efficient fusion which by fusing different intermediate layers, each modality learns complementary contextual information from the others. And Burchi et al. introduced the AVEC model~\cite{burchi2023audio}. This model uses Efficient Conformer blocks~\cite{burchi2021efficient} to reduce parameters while preserving learning capacity by incorporating the Inter-CTC module. However, like many AVSR models, AVEC directly concatenates audio and visual features, which forces the model to learn the hierarchical relationship between these modalities, unintentionally increasing its learning burden.

Inspired by the above models, we propose a novel asymmetric architecture that {\it prioritizes the audio modality while treating the visual modality as supplementary} for the efficient AVSR. This design allows for more efficient integration of multi-modal information. Our primary contribution is the introduction of a new AVSR model architecture that uses only a small amount of the Conformer modules to extract visual features and fully integrate them into the audio features. Central to this architecture is the {\it Dual Conformer Interaction Module (DCIM)}, which significantly enhances cross-modal information exchange between audio and visual inputs. Additionally, we developed a pre-training method that further improves performance. We use Inter-CTC loss~\cite{9414594} in the DCIM module to restrict the learning process of features. Our model achieves a 14\% relative reduction in parameters while a 13\% relative reduction in Word Error Rate (WER) compared to the baselines on LRS2~\cite{Afouras18c} and LRS3~\cite{afouras2018lrs3}. We conducted an ablation study to demonstrate the effectiveness of each module we introduced. The impact of our work lies in its potential to set a new standard for efficient AVSR models, offering a promising direction for future research in this domain.

\begin{figure*}[htbp]
\centerline{\includegraphics[scale=0.033]{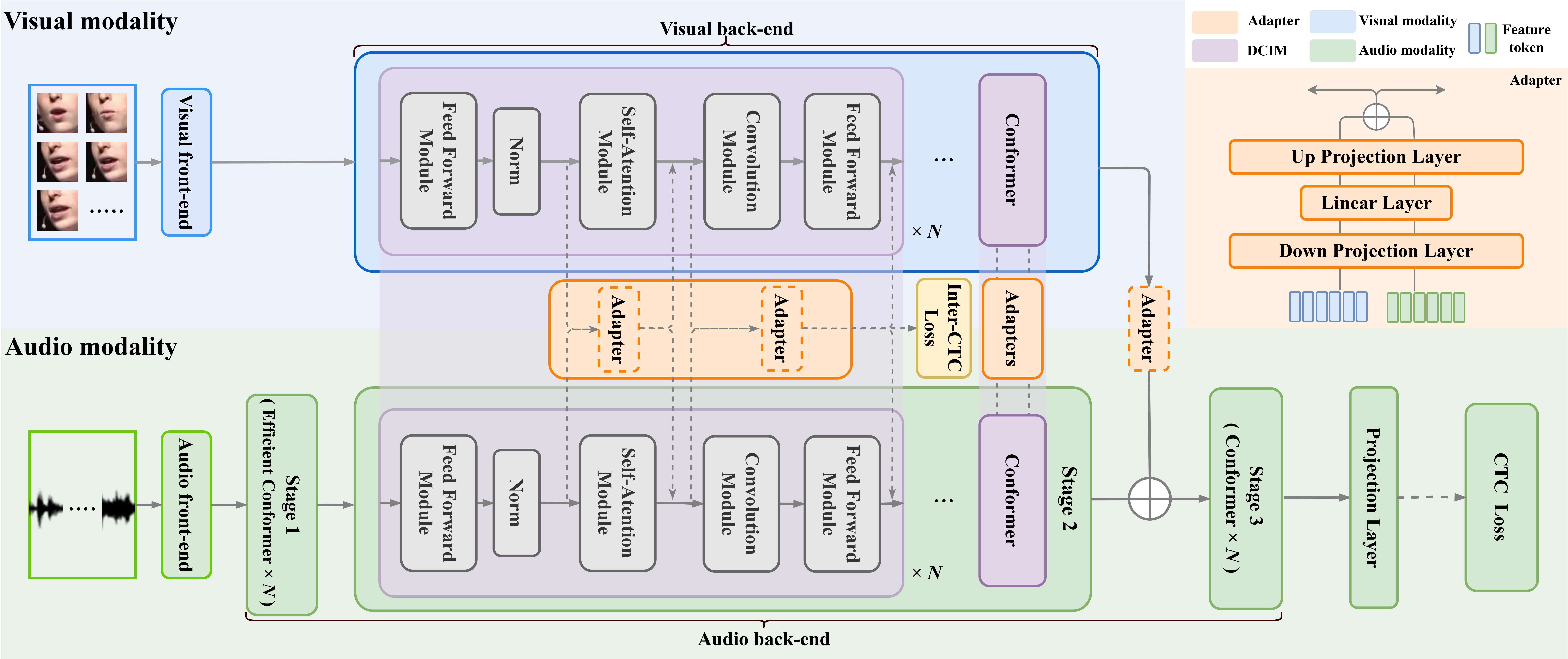}}
\vspace{-3mm}
 \caption{The overall architecture of the DCIM-AVSR (Dual-Mode Conformer Interaction Model for Audio-Visual Speech Recognition) and the adapter module illustrates the mechanism and flow of cross-modal information interaction. }
\vspace{-3mm}
\label{fig1}
\end{figure*}
\section{METHOD}

\subsection{Overall Architecture}\label{AA}

As shown in Fig.\ref{fig1}, our model is composed of five main components: the visual front-end, the audio front-end, the visual back-end, the audio back-end and the {\it Dual Conformer Interaction Module}(DCIM). The {\it visual front-end} consists of a 3D convolutional layer followed by four layers of 2D ResNet-18~\cite{he2016deep} and a Global Average Pooling layer. The {\it audio front-end} converts the audio into a Mel-spectrogram, which is then processed through 2D convolutional layers. The {\it audio and visual back-end} are comprised of multiple layers of Conformer blocks. To enhance the model's efficiency, we design the audio and visual layers asymmetrically, with one stage for visual and three stages for audio. The first audio stage uses 5 Efficient Conformer\cite{burchi2021efficient}, while the remaining two stages use the 5 and 4 layers standard Conformer\cite{gulati20_interspeech}. The middle audio stage and the visual stage form the Dual Conformer Interaction Module (DCIM), facilitating information exchange between the two modalities. 
Visual back-end consists of 5 layers of Conformer modules, and to maintain model performance, the features from each layer of the visual modality are integrated with the audio modality through the DCIM module.  Finally, the output of the visual back-end is filtered through an adapter and added to the audio features as supplementary information.
The final output of the audio branch is used as the final result of the AVSR model. Specifically, our training strategies are detailed in \ref{setion}.

The above design ensures that the model pays more attention on how to learn the characteristics of the main modality and compensates for the information of the main modality with auxiliary modality.
In other words, it's a more efficient strategy that reduces redundant computing.      We will demonstrate this in experiments. 

\subsection{Dual Conformer Interaction Module}
\label{DCIM}

The Dual Conformer Interaction Module is designed for distribute the fusion task between audio and visual to each DCIM, facilitating efficient cross-modal information exchange.
As shown in the Fig.\ref{fig1}, the Dual Conformer Interaction Module proposed in this paper consists of two Conformer modules and two adapter modules. The two Conformer are from the back-end of visual and audio respectively. And adapter modules are inserted at various points within the Conformer' structure. We use the output of the second adapter in even DCIM layers to calculate Inter-CTC loss~\cite{9414594}. 

The output to the \emph{i}-th DCIM can be described as,
\begin{equation}
(x_{v}^{i+1}, x_{a}^{i+1}) = \mathit{F^{i}}(x_{v}^{i}, x_{a}^{i})   \quad i = 1, 2, 3,\cdots,N
\label{eq1}
\end{equation}

where $\mathit{F^i}$ represents the $\mathit{i}$-th DCIM module, and $\mathit{x^{i}}$ represents the \emph{i}-th layer's input of the corresponding modality.

\subsection{Adapter in DCIM}

\begin{figure*}[htbp]
\centerline{\includegraphics[scale = 0.12]{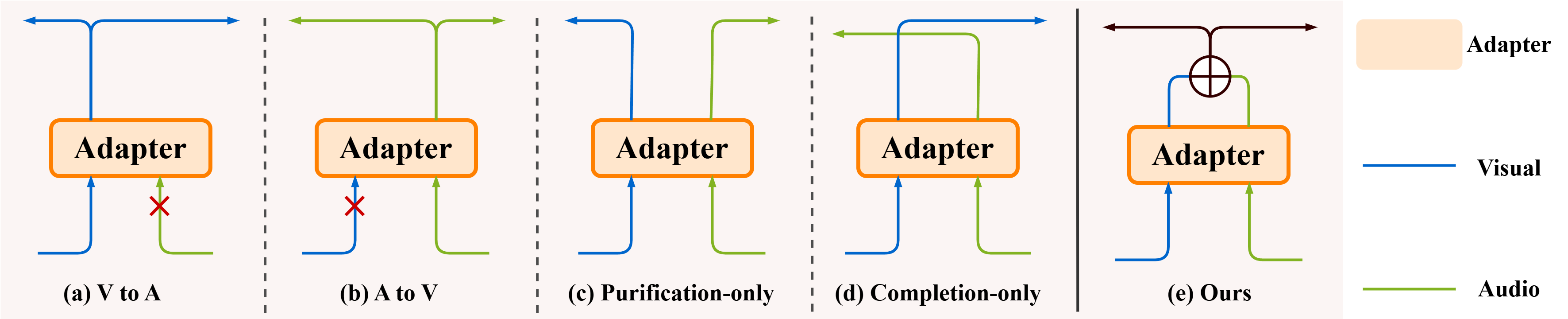}}

\vspace{-3mm}
\caption{The detailed comparison of the different variants of adapter, focusing on their functionality.}
\label{fig2}
\vspace{-3mm}
\end{figure*}

In previous work~\cite{guo2023npu, wang2024mlca, berghi2024fusion}, various cross-modal fusion modules were proposed, with {\it Cross-modal  Attention}(CA)~\cite{guo2023npu} being the most common.  It functions by exchanging the K and V matrices of the two modalities' attention module.  But the audio signal changes rapidly, whereas lip movements in the video are relatively slow and smooth.  As a result, the features may still differ at each point in time.  This discrepancy increases the burden on the cross-attention module, which must not only learn the recognition task for each modality but also manage the balance of feature relations between modalities, making it an inefficient fusion method. 

In order to solve the above problems, inspired by Bi-directional Adapter~\cite{cao2024bi}, we propose a kind of efficient adapter that suitable for Conformer architecture. By introducing a specialized adapter module between the two modalities, the learning process is divided into two parts.   The Conformer focuses solely on processing the features of each modality individually, while the adapter layer learns to selectively enhance or suppress certain features before feeding them back to both modalities.   This approach enables more efficient cross-modal interactions and improves overall performance. The specific architecture of adapter is shown in the Fig.\ref{fig1}.

The adapter consists of three linear projection layers. First, the input visual or audio features are reduced in dimension and processed through a linear projection layer. These features are then projected back to their original dimension and used as informational features for the respective modality.

We add two adapter modules to the DICM module. In Conformer\cite{gulati20_interspeech}, the {\it Self-Attention} and {\it Convolution} modules play crucial roles: {\it Self-Attention} captures global dependencies within a sequence, while {\it Convolution} captures local dependencies. The combination of these two mechanisms makes Conformer one of the most powerful backbone in the field of speech processing. And we took a deeper look into the two modules. Based on the functions of the {\it Self-Attention}  and {\it Convolution} modules, we added adapter modules at the corresponding locations. Through two adapter modules, global and local information from different modalities can interact and merge more effectively. This design enhances the model's ability to learn subtle dependencies between the two modalities, thereby improving overall performance.

 The proposed adapter serves two primary functions within the model as show in Fig.\ref{fig2}(e): information {\it completion} and {\it purification}. {\it  Purification} involves transmitting information within a modality. 
 This allows the model to consistently learn valuable feature information. {\it  Completion} refers to processing features from different modalities through the adapter, which  enrich the information within each modality. Functions of the adapter can be described as,
\begin{equation}
\begin{aligned}
I_{m}^{'} = Ada(X_{m})& + Attention(X_{m}) + Ada(X_{\widetilde{m}}),\\
X_{m} &= Norm(FFM(x_{m}))
\label{eq2}
\end{aligned}
\end{equation}
\begin{equation}
\begin{aligned}
I_{m}^{''} = Ada(I_{m}^{'}) + FFM(Conv(I_{m}^{'})) + Ada(I_{\widetilde{m}}^{'})
\label{eq3}
\end{aligned}
\end{equation}

where $\mathit{I}$ is the intermediate feature after the adapter module. And $\mathit{m}$ represents either modality, $\mathit{\widetilde{m}}$ represents the other modality. $\mathit{FFM}$, $\mathit{Conv}$ and $\mathit{Norm}$ stand for feed-forward module, self-attention module and norm in Conformer block.

This dual-modal adapter enables more efficient fusion of audio and visual modalities, allowing the model to learn information from both modalities in each DCIM and thereby improving the model's representation capability.

\subsection{Training strategy}
\label{setion}

We divide the training into three stages: ASR pre-training, VSR pre-training, and AVSR fine-tuning. 
The pre-training used here is not generalized unsupervised training but is intended to improve the model’s ability to initialize feature extraction weights. First, the ASR and VSR models are trained separately, with both models having the same number of layers as the final AVSR model. It is important to note that the training here is not expected to produce optimal results; rather, the purpose of pre-training is to enable the model to learn how to extract features from the raw data. The resulting weights of the audio and visual branches are then used to initialize the AVSR model. 

\section{Experiments}

\subsection{Datasets and Pre-processing}

We utilized three public datasets: LRW (Lip Reading in the Wild)~\cite{chung2017lip}, LRS2 (Lip Reading Sentences 2)~\cite{Afouras18c} and LRS3 (Lip Reading Sentences 3)~\cite{afouras2018lrs3}. The LRW dataset was used for pre-training the visual front-end, while LRS2 and LRS3 served as the training and test sets for our model. For visual data, we followed the method outlined in previous work~\cite{ma2021end} by cropping the lip region to a size of 96×96 pixels, converting it to grayscale, and normalizing it. During training, we applied data augmentation techniques such as random cropping, flipping, and rotation. For audio data, we first converted it into a Mel-spectrogram and then applied SpecAugment~\cite{park2020specaugment} for data augmentation.
\subsection{Implementation Details}
\label{setup}

Our model is divided into three components: ASR, VSR, and AVSR. Experiments were conducted on an A100 GPU, with the parameter counts for the ASR, VSR, and AVSR models being 22M, 29M, and 53M, respectively. The audio branch has dimensions [180, 256, 360], changing every five layers, while the visual branch has dimensions [256, 360]. All branches use 4 attention heads, a kernel size of 15, and a vocabulary size of 256. The adapters used in the DCIM have dimensions [256, 180, 256] when Conformer-dim=256 and [360, 256, 360] when Conformer-dim=360.

For the VSR model, we first pre-trained a visual front-end using the LRW dataset~\cite{chung2017lip} and then trained a 5-layers VSR model based on this visual front-end. All three models utilized the Adam optimizer with a warmup learning rate scheduler. The ASR model was trained for 100 epochs with a batch size of 64, the VSR model for 30 epochs with a batch size of 32, and after pre-training the ASR and VSR models, the AVSR model was trained for 20 epochs with a batch size of 32. Additionally, we trained the AVSR model directly for 80 epochs with the same architecture and batch size. In terms of training efficiency, ASR/VSR pre-training followed by AVSR parameter-efficient tuning significantly reduces training costs. And we used a language model for decoding: a GPT-small model trained on LibriSpeech and fine-tuned on LRS2 and LRS3.
\section{Result and Analysis}

\subsection{Comparison to the state-of-the-art}
We trained the DCIM-AVSR model both from scratch and using the pre-training approach. And to demonstrate the efficiency of our models, we compare them with the AVEC~\cite{burchi2023audio} model, which has achieved state-of-the-art results on LRW~\cite{chung2017lip}, LRS2~\cite{Afouras18c} and LRS3~\cite{afouras2018lrs3} (without using additional datasets). As shown in the Table~\ref{tab2}, both methods yield improved results. Furthermore, to verify the effectiveness of the DCIM module, we integrated it into the AVEC model, which also outperformed the baseline. At the same time, we also added the CA~\cite{guo2023npu} module to AVEC as a contrast. Experiments show that the CA module does not bring significant performance improvement compared to the DICM we proposed. Meanwhile, the asymmetric model architecture and the training method based on DICM have achieved impressive performance.

Additionally, compared to other AVSR models. 
For example, while the AV-Hubert large model uses 325M parameters to achieve a WER of 1.4\% on LRS3, our model achieves a comparable WER with only 53M parameters. 
Our efficient DCIM-AVSR model requires far fewer parameters and computational resources than other existing AVSR models. Although there may be a slight gap in performance, it can still operate effectively in resource-constrained environments due to its significant reduction in parameter count and training data requirements.

\begin{table}[t]
    \centering
    \caption{Comparison of WER(\%) on LRS2/LRS3 with other AVSR models}
    \renewcommand\arraystretch{1.2}
    \setlength{\tabcolsep}{2.5mm}
    \vspace{-2mm}
    \begin{tabular}{l|c c c c}
    \hline
        Method & Params & Datasets(hours) & LRS2 & LRS3  \\ 
        \hline \hline
        AV-Hubert large\cite{DBLP:conf/iclr/ShiHLM22} & 325M & 2192  & / & 1.4  \\ \hline
        BAVen\cite{haliassos2023jointly} & 328M & 1759  & / & 1.4  \\ \hline
        Auto-AVSR\cite{ma2023auto} & 425M & 3448  & 1.5 & 0.9  \\ \hline
        LP Conformer\cite{chang2024Conformer} & 570M & 100k & / &0.9  \\ \hline
        AV-Hubert base~\cite{DBLP:conf/iclr/ShiHLM22} & 103M & 2192 & / & 1.8  \\ \hline \hline
        AVEC~\cite{burchi2023audio} & 61M & 818 & 2.31 & 1.82  \\
        \quad +CA~\cite{guo2023npu} & 61M & 818 & 2.37 & 1.74  \\ 
        \quad +DCIM & 62M & 818 & 2.15 & 1.70  \\ \hline
        Ours & 53M & 818 & 2.04 & 1.68  \\ 
        \quad +Pre-training & 53M & 818 &   \pmb{1.95} &   \pmb{1.62} \\ \hline

    \end{tabular}
    \vspace{-6mm}
    \label{tab2}
\end{table}

\subsection{Effect of Noise robustness}

\begin{figure}[htbp]
    \centerline{\includegraphics[scale = 0.185]{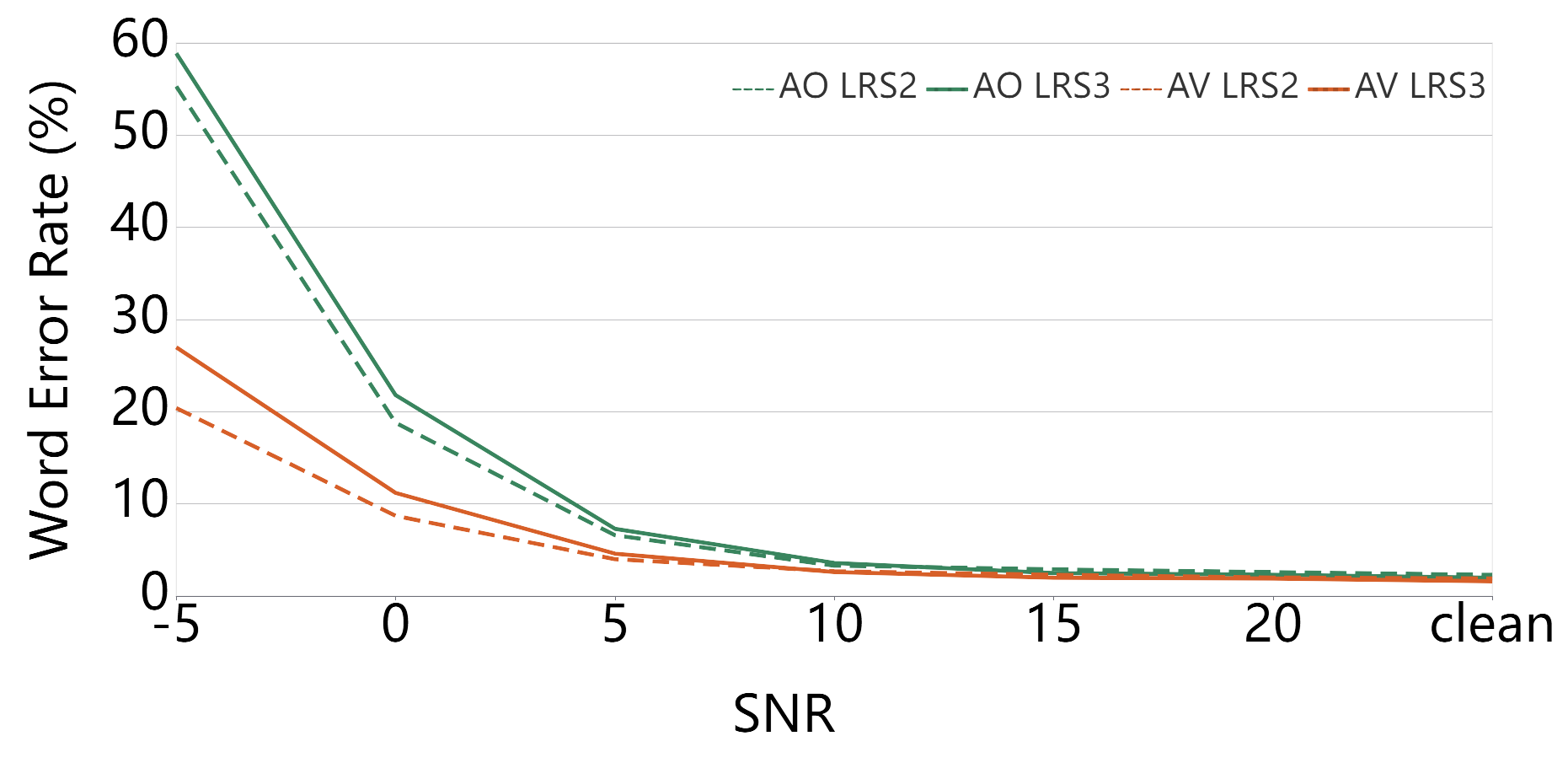}}
    
    \vspace{-3mm}
    \caption{WER (\%)  Comparison on LRS2/LRS3 Under Various Conditions.}
    \label{tab3}
    \vspace{-3mm}
\end{figure}

Robustness in complex situations is crucial for AVSR models. To test this, we introduced white noise from the Noisex-92~\cite{VARGA1993247} dataset. We compare an audio-only model with our audio-visual (AV) model. The noise has SNR in [-5, 0, 5, 10, 15, 20]. As shown in Fig.\ref{tab3},  our model has superior robustness across various acoustic environments.


\subsection{Ablation study}

To further demonstrate the advantages of the DCIM module, we analyze its construction under various configurations. As shown in Table~\ref{tab4}, all experiments were conducted using consistent settings as~\ref{setup}. The DCIM module processes features from both audio and visual modalities for completion and purification, which significantly enhances performance. Our experiments reveal that processing features from only one modality (as shown in Fig.\ref{fig2}(a)(b)) leads to poorer performance.  
This reduction in performance occurs because one-way processing limits the model's ability to learn from both modalities, diminishing its effectiveness in extracting common features' essential for accurate recognition.
Additionally, we verify the importance of both modal information completion and purification(as shown in Fig.\ref{fig2}(c)(d)) within DCIM. When the model performs only purification, the WER increases to 2.46\% and 1.87\% on LRS2 and LRS3, respectively. This is because visual information is integrated with audio features only at the final layer, which neither reduces the learning cost nor improves model performance.
Furthermore, we explored the impact of the number of DCIM layers on model performance. Comparing the use of DCIM in only the last two layers versus all five layers showed that incorporating all five layers yields better performance. This indicates that increasing the number of DCIM layers contributes to enhanced accuracy.
\begin{table}[t]
    \centering
    \caption{Results (WER \%) of DCIM with different structures.}
    \renewcommand\arraystretch{1.2}
    \setlength{\tabcolsep}{2.5mm}
    \vspace{-2mm}
    \begin{tabular}{l|ccccc}
    \hline
        Type & All Layers & Purification & Completion & LRS2 & LRS3  \\ \hline\hline
        V to A & \ding{51} & \ding{51} & \ding{51} & 2.19 & 1.70  \\ \hline
        A to V & \ding{51} & \ding{51} & \ding{51} & 2.12 & 1.73  \\ \hline
        Dual & \ding{55} & \ding{51} & \ding{51} & 2.21 & 1.75  \\ \hline
        Dual & \ding{51} & \ding{55} & \ding{51} & 2.05 & 1.66  \\ \hline
        Dual & \ding{51} & \ding{51} & \ding{55} & 2.46 & 1.87  \\ \hline
        Dual & \ding{51} & \ding{51} & \ding{51} & \pmb{1.95} & \pmb{1.62}  \\ \hline
        
    \end{tabular}
    \vspace{-6mm}
    \label{tab4}
    
\end{table}

\section{Conclusion}

In this paper, we introduce an asymmetric DCIM-AVSR model, where the visual modality serves as the auxiliary modality and the audio modality as the primary modality. We also propose a Dual Conformer Interaction Module (DCIM) to enhance multi-modal interaction. The model distributes the information fusion process across each DCIM module. Additionally, we introduce a pre-training strategy that maximizes the benefits of the pre-trained model, reducing learning and training costs while improving performance. Our experiments show that the DCIM-AVSR model achieves better performance compared to conventional AVSR models with great robustness.

\clearpage
\balance
\bibliographystyle{ieeetr}
\bibliography{reference}

\vspace{12pt}

\end{document}